# Lattice and Magnetic structures of PrFeAsO, PrFeAsO$_{0.85}$F$_{0.15}$ and PrFeAsO$_{0.85}$

Jun Zhao[1], Q. Huang[2], Clarina de la Cruz[1,3], J. W. Lynn[2], M. D. Lumsden[3], Z. A. Ren[4], Jie Yang[4], Xiaolin Shen[4], Xiaoli Dong[4], Zhongxian Zhao[4], and Pengcheng Dai[1,3]

[1]Department of Physics and Astronomy, The University of Tennessee, Knoxville, Tennessee 37996-1200

[2] NIST Center for Neutron Research, National Institute of Standards and Technology, Gaithersburg, Maryland 20899-6012

[3]Neutron Scattering Science Division, Oak Ridge National Laboratory, Oak Ridge, Tennessee 37831

[4]Beijing National Laboratory for Condensed Matter Physics, Institute of Physics, Chinese Academy of Sciences, Beijing 100080, China

Abstract

We use powder neutron diffraction to study the spin and lattice structures of polycrystalline samples of nonsuperconducting PrFeAsO and superconducting PrFeAsO$_{0.85}$F$_{0.15}$ and PrFeAsO$_{0.85}$. We find that PrFeAsO exhibits an abrupt structural phase transitions at 153 K, followed by static long range antiferromagnetic order at 127 K. Both the structural distortion and magnetic order are identical to other rare-earth oxypnictides. Electron-doping the system with either Fluorine or oxygen deficiency suppresses the structural distortion and static long range antiferromagnetic order, therefore placing these materials into the same class of FeAs-based superconductors.

**Introduction**

The recent discovery of superconductivity in the rare-earth (R) iron-based superconductors $RFeAsO_{1-x}F_x$ [1-4] and $A_{1-x}K_xFe_2As_2$ (A=Ba, Sr) [5,6] has generated enormous interest because these materials are the first non-copper oxide superconductors with $T_c$ up to 55 K. All of the parent compounds (RFeAsO, R=La,Ce,Nd; $AFe_2As_2$, A=Ba, Sr,) of the iron-based superconductors investigated so far display a similar antiferromagnetic phase transition accompanying a tetragonal to orthorhombic structural distortion on cooling from 250K to 100K [7-18]. Upon doping with fluorine (F), the long range antiferromagnetic order is gradually suppressed before superconductivity appears, indicating a remarkably similar electron phase diagram as the copper oxide superconductors [10]. Although the lattice structure and magnetic properties of $RFeAsO_{1-x}F_x$ (R=La,Ce,Nd) are similar, the maximum superconducting temperature $T_c$ of doped $RFeAsO_{1-x}F_x$ is dramatically different, increasing from 26K to 50K when R changes from the non-magnetic element La to magnetic elements Ce, Nd and Pr. The sensitivity of the superconductivity to rare earth substitution is really surprising given the separate FeAs layers, and completely different from copper oxide superconductors in which the superconductivity is only sensitive to the element substitution within the $CuO_2$ plane (except for the case of $PrBa_2Cu_3O_{6+x}$ and $CeBa_2Cu_3O_{6+x}$). In order to understand how the rare earth substitution controls the $T_c$, it is important to study how the rare earth substitution affects the lattice and magnetic structure and the resulting electron band structures.

In this paper, we present neutron scattering studies of the parent compound PrFeAsO and its superconducting counterpart $PrFeAsO_{1-x}F_x$, which possesses the highest $T_c$ in the FeAs-based superconductor series [19].  Furthermore, since superconductivity in PrFeAsO can also be induced by forming oxygen vacancies, comparing the structure and magnetic properties of the F-doped and oxygen deficient samples should provide some clues to understanding the role of doping in the FeAs-based class of superconductors.  Here we investigate the structure and magnetic properties of the parent compound PrFeAsO and its superconducting counterparts PrFeAsO$_{0.85}$F$_{0.15}$ ($T_c$ = 52 K) and PrFeAsO$_{0.85}$ ($T_c$ = 52 K) by elastic neutron scattering.  We find that PrFeAsO undergoes a structural distortion from tetragonal to orthorhombic symmetry near 153 K, accompanied by a magnetic transition to commensurate antiferromagnetic order of the Fe spins at ~127 K.  These results, taken together with the observations of magnetic order in all the other systems which have been investigated to date [8-18], demonstrate that the antiferromagnetic order is universal for the parent compounds of the FeAs-based superconductors.  Upon 15% F doping, both the structural distortion and magnetic order are suppressed, identical to the other FeAs-based superconductors.  We also find that the structural distortion and magnetic order are suppressed in the oxygen-deficient superconducting PrFeAsO$_{0.85}$ sample.  Thus, removing oxygen from PrFeAsO has the same impact on the structural and magnetic properties as doping F in the system.

**Experimental Results and Discussion**

We have employed neutron diffraction to study the structural and magnetic order in polycrystalline samples of PrFeAsO, PrFeAsO$_{0.85}$F$_{0.15}$ ($T_c$ = 52K) and PrFeAsO$_{0.85}$ ($T_c$ = 52 K).  The samples were synthesized by a high pressure method as described in ref 19.  Our neutron scattering experiments were carried out on the BT-1 powder diffractometer at the NIST Center for Neutron Research (NCNR), using the Ge(3,1,1) monochromator with an incident beam wavelength of $\lambda$ = 2.0785 Å.  The collimations before and after the monochromator and after the sample were 15′, 20′, and 7′ full-width-at-half-maximum (FWHM), respectively.  Magnetic order parameters were taken on the HB-3 thermal triple-axis spectrometer at High Flux Isotope Reactor, Oak Ridge National Laboratory, with an incident beam wavelength $\lambda$=2.36Å with pyrolytic graphite (PG) (0,0,2) as monochromator and PG filters.  Collimations in these configurations were coarse (typically ~40-50′), for intensity reasons.

Our high-resolution measurements on BT-1 show that the high temperature structure (175 K) in PrFeAsO can be well described by the expected tetragonal structure of space group *P4/nmm* (Figure 1c). The refined structural parameters are listed in Table 1a. Figure 1d shows the low temperature (5 K) diffraction pattern and refinement profiles for PrFeAsO, which can be described with the orthorhombic structure of space group *Cmma*. The orthorhombic distortion splits the $(220)_T$ of the tetragonal structure into two peaks, $(400)_O$ and $(040)_O$ in the orthorhombic structure, as shown in the inset of Figure 2a. The insets in figure 1c and 1d show the details of the diffraction pattern for 2θ between 20 to 37 degrees, where most of the observable magnetic peaks are located. We can clearly see several magnetic peaks at 5 K, which can be simply indexed with the expected commensurate magnetic structure. These peaks are absent in the 175K diffraction pattern, indicating that we are in the paramagnetic state at this temperature. Refinements using the GSAS program give excellent fits for the low temperature diffraction pattern, where the magnetic peaks are well accounted for by the combined Pr and Fe antiferromagnetic order as shown in Figure 1a and 1b. The Fe magnetic unit cell can be indexed as $\sqrt{2}a_N \times \sqrt{2}b_N \times c_N$, which is exactly the same as for CeFeAsO [10]. The Fe spins order antiferromagnetically along the orthorhombic *a* axis and ferromagnetically along the *b* and *c* axis, with the moment direction along the *a* axis. The measured static ordered Fe moment is 0.48(9) $\mu_B$ at 5K, where numbers in parentheses indicate one standard deviation statistical uncertainty in the last decimal place and $\mu_B$ denotes the Bohr magneton. The Pr spins order antiferromagnetically as shown in Figure 1a. The static ordered Pr moment is 0.84(4) $\mu_B$ at 5K.

Figure 2 plots the order parameter data for the structural and magnetic phase transitions. The onset of the structural transition is indicated by the initial drop in the $(220)_T$ peak intensity with temperature, which is observed to be around 153 K (Figure 2a). Figure 2b reveals that the Pr Néel temperature is about 14 K, while the Fe Néel temperature is about 127 K. Compared to the undoped PrFeAsO system, there is no observable orthorhombic structure distortion in the PrFeAsO$_{0.85}$F$_{0.15}$ down to 5 K (Figure 3a). The tetragonal *P4/nmm* structure can describe the diffraction pattern very well, as is the case for all the other highly doped FeAs-based superconductors. The oxygen-deficient PrFeAsO$_{0.85}$ sample (Fig. 3b) also has no orthorhombic structural distortion down to 5 K, and the refined structural parameters are essentially the same as for the F doped sample (Table 1b, 1c). In addition, neither PrFeAsO$_{0.85}$F$_{0.15}$ nor PrFeAsO$_{0.85}$ has any observable magnetic order at 5 K, suggesting that antiferromagnetic order is directly competing with superconductivity.

## Conclusions

To summarize, we have carried out detailed neutron scattering studies of the magnetic and nuclear structures of the FeAs-based superconductors PrFeAsO$_{0.85}$F$_{0.15}$ ($T_c$ = 52 K), and PrFeAsO$_{0.85}$ ($T_c$ = 52 K), along with their parent compound PrFeAsO. Very similar to the other parent compounds of FeAs-based superconductors, PrFeAsO has a simple stripe-type antiferromagnetic structure of the iron spins, with a Néel temperature of 127K and an ordered moment of 0.48(9) $\mu_B$. The magnetic moments on the Pr sites are also antiferromagnetically ordered below 14 K, similar to the parent compounds of the other rare earth FeAs-based superconductors such as CeFeAsO [10] and NdFeAsO [12]. The iron magnetic order occurs below the transition from the high temperature tetragonal phase to the low temperature orthorhombic phase of the parent compound that occurs around 153K. The structural distortion and iron antiferromagnetic order are suppressed completely in the optimally doped superconducting samples, regardless of whether the superconducting state is achieved by F doping or oxygen vacancies, and the two types of doping yield very similar crystallographic structures.

## Acknowledgements

This work is supported by the US National Science Foundation through DMR-0756568, by the US Department of Energy, Division of Materials Science, Basic Energy Sciences, through DOE DE-FG02-05ER46202. This work is also supported in part by the US Department of Energy, Division of Scientific User Facilities, Basic Energy Sciences. The work at the Institute of Physics, Chinese Academy of Sciences, is supported by the Note added: Upon finishing the present paper, we became aware of a similar neutron powder diffraction work on PrFeAsO and PrFeAsO$_{1-x}$F$_x$ (ref. 20).

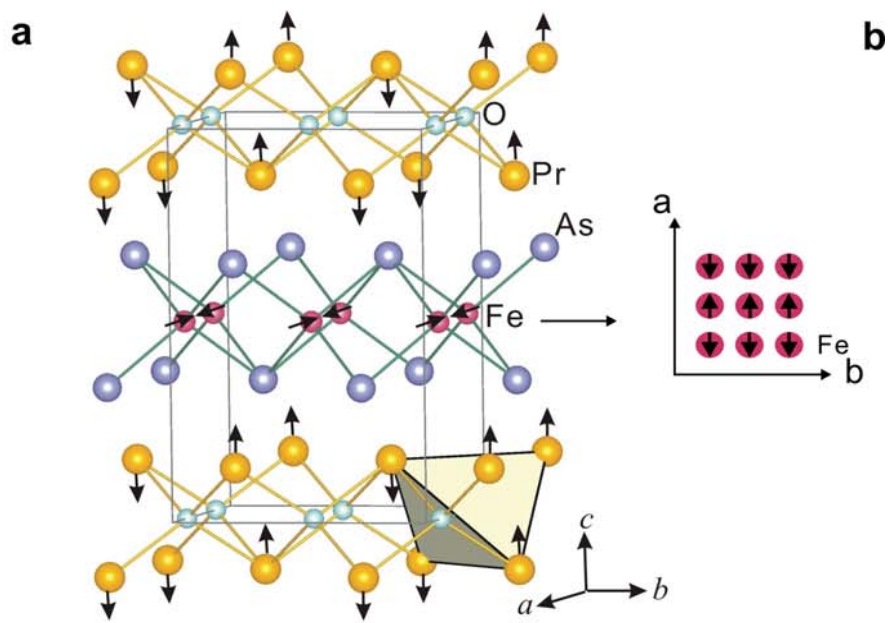
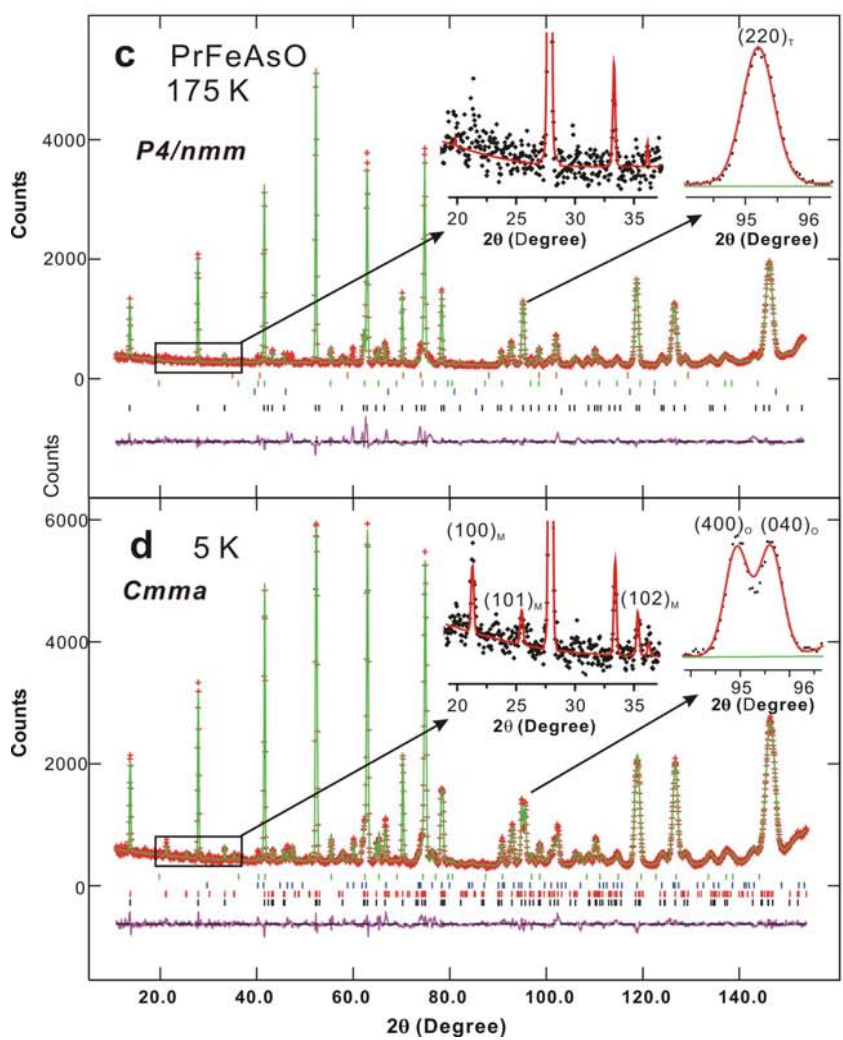

**Figure 1** (color online) Lattice and magnetic structures for Fe and Pr in undoped PrFeAsO. **a**) The three dimensional antiferromagnetic structures of Fe and Pr as determined from the refinements of our neutron diffraction data. **b**) The magnetic structure of Fe in the FeAs plane. **c**) Observed (crosses) and calculated (solid line) neutron diffraction intensities of PrFeAsO at 175 K, in the tetragonal structure with space group *P4/nmm*. The inset shows the detailed data for 18°<2θ<38°, where most of the observable magnetic peaks are located. 2θ is the diffraction angle, and the short vertical lines show the Bragg peak positions. No magnetic peaks are observed at 175K. The (purple) trace indicates the intensity difference between the observed and calculated structures. **d**) Diffraction data at 5 K, fit with the orthorhombic structure of space group *Cmma*. The inset plots the detailed data in 18°<2θ<38° showing three indexed magnetic peaks at 5K, along with the observed splitting of the structural peak. The magnetic peaks are accounted for by the combined contributions of Fe and Pr. The (1,0,0) peak vanishes completely above the Pr Néel temperature of 14 K, while the (1,0,1) and (1,0,2) peaks persist above 14 K and vanish at 127 K as shown in Figure 2b.

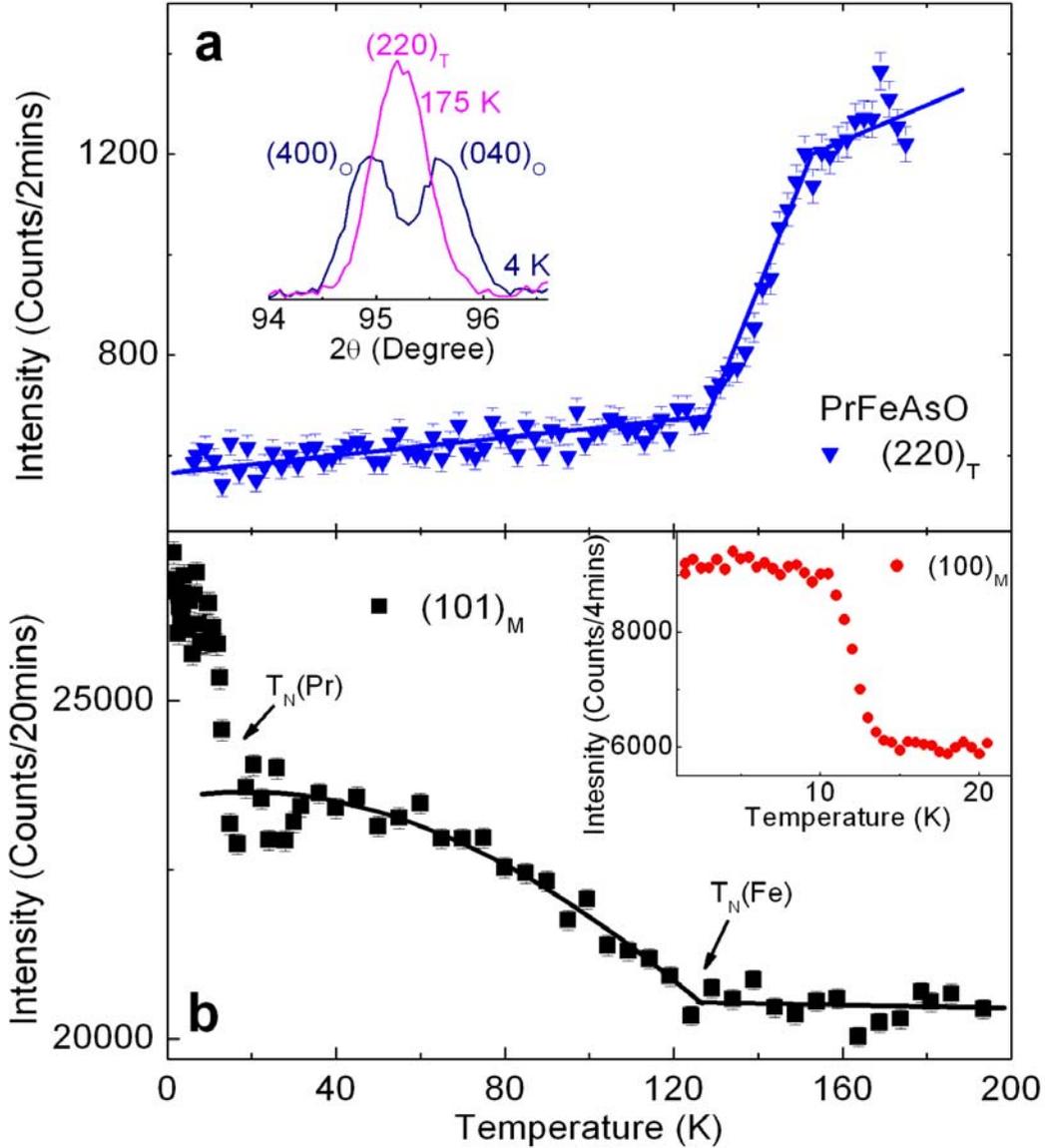

**Figure 2** (color online) Temperature dependence of $(2,2,0)_T$ nuclear Bragg peak and magnetic $(1,0,0)_M$ and $(1,0,1)_M$ peaks. The data in a) and b) are collected on BT-1 and HB-3, respectively. **a)** Temperature dependence of the $(2, 2, 0)_T$ (T denotes tetragonal) nuclear Bragg peak showing the onset of the structure phase transition is about 153 K. The inset shows the $(2, 2, 0)_T$ reflection above and below the transition temperature. **b)**

Temperature dependence of the order parameter for the $(1,0,1)_M$ (M denotes Magnetic) magnetic Bragg peak. The large increase of the intensity below 14 K is due to Pr ordering, as confirmed by the temperature dependence of the $(0, 0, 1)_M$ magnetic Bragg peak, which has only an intensity contribution from Pr. The intensity of the $(1,0,1)_M$ peak vanishes at the Néel temperature of 127 K for the iron spin ordering.

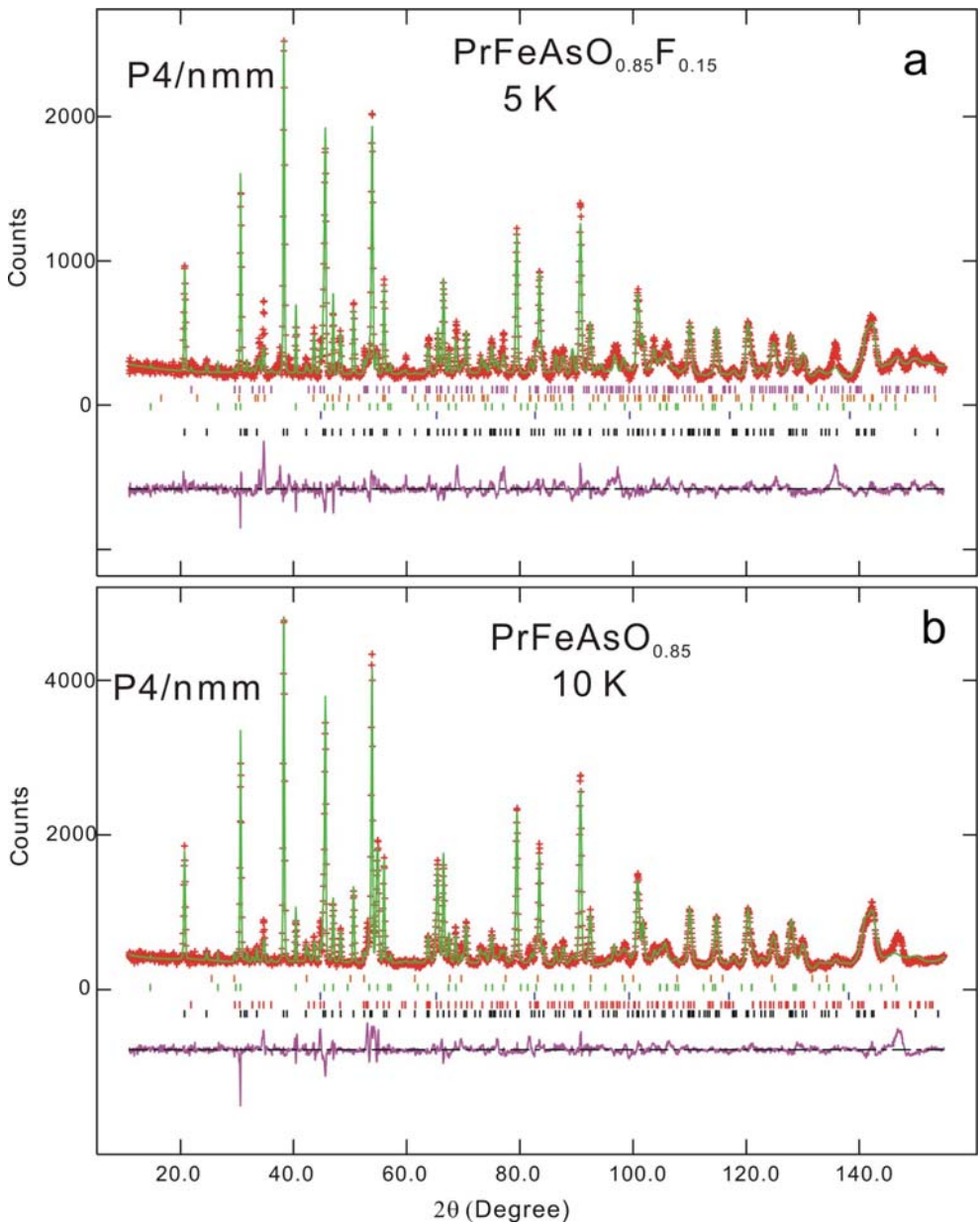

**Figure 3** (color online) Structural diffraction data for the PrFeAsO$_{0.85}$F$_{0.15}$ and PrFeAsO$_{0.85}$ superconducting samples. The data were collected on the BT-1 diffractometer. **a)** Observed (crosses) and calculated (solid line) neutron diffraction intensities of PrFeAsO$_{0.85}$F$_{0.15}$ at 5 K for the tetragonal structure (space group *P4/nmm*). The short vertical lines show the Bragg peak positions. The (purple/grey) trace indicates the intensity difference between the observed and calculated structures. **b)** Observed (crosses) and calculated (solid line) neutron diffraction intensities of PrFeAsO$_{0.85}$ at 5 K, refined with the tetragonal space group *P4/nmm*.

**Table 1a**. Refined structural parameters of PrFeAsO$_{1-x}$F$_x$ with $x = 0$ at 175 K, $x = 0.15$ at 5 K and PrFeAsO$_{0.85}$ at 5 K. Space group: *P4/nmm*. PrFeAsO, $a = 3.97716(5)$, $c=8.6057(2)$ Å; PrFeAsO$_{0.85}$F$_{0.15}$, $a = 3.9700(1)$, $c = 8.5331(4)$ Å; PrFeAsO$_{0.85}$, $a = 3.9686(1)$, $c = 8.5365(3)$ Å

| Atom | site | $x$ | $y$ | $z$(PrFeAsO) | $z$ (PrFeAsO$_{0.85}$F$_{0.15}$) | $z$ (PrFeAsO$_{0.85}$) |
|---|---|---|---|---|---|---|
| Pr | 2c | ¼ | ¼ | 0.1397(6) | 0.1504(1) | 0.1450(7) |
| Fe | 2b | ¾ | ¼ | ½ | ½ | ½ |
| As | 2c | ¼ | ¼ | 0.6559(4) | 0.6548(5) | 0.6546(5) |
| O/F | 2a | ¾ | ¼ | 0 | 0 | 0 |

PrFeAsO, $Rp =4.55\%$, $wRp = 5.8\%$, $\chi^2= 1.387$;

PrFeAsO$_{0.85}$F$_{0.15}$, $Rp =8.24\%$, $wRp = 10.62\%$, $\chi^2 = 3.635$.

PrFeAsO$_{0.85}$, $Rp =6.99\%$, $wRp = 9.23\%$, $\chi^2 = 4.652$

**Table 1b**. Refined structural parameters of PrFeAsO at 5 K. Space group: *Cmma*. $a = 5.6374(1)$, $b = 5.6063(1)$, $c = 8.5966(2)$ Å

| Atom | site | $x$ | $y$ | $z$ |
|---|---|---|---|---|
| Pr | 4g | 0 | ¼ | 0.1385(5) |
| Fe | 4b | ¼ | 0 | ½ |

| As | 4g | 0 | ¼ | 0.6565(3) |
| O  | 4a | ¼ | 0 | 0 |

$Rp$ =4.22%, $wRp$ = 5.73%, $\chi^2$= 2.180;